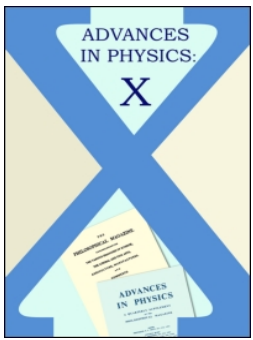

# Advances in Physics: X

# Phonon coherence and its effect on thermal conductivity of nanostructures

**Guofeng Xie, Ding Ding & Gang Zhang**

Submit your article to this journal 🗗

View Crossmark data 🗗

Taylor & Francis
Taylor & Francis Group

# Phonon coherence and its effect on thermal conductivity of nanostructures

Guofeng Xie[a], Ding Ding[b] and Gang Zhang[c]

[a]School of Materials Science and Engineering, Hunan University of Science and Technology, Xiangtan, China; [b]Singapore Institute of Manufacturing Technology, Singapore, Singapore; [c]Institute of High Performance Computing, Singapore, Singapore

## ABSTRACT

The concept of coherence is one of the fundamental phenomena in electronics and optics. In addition to electron and photon, phonon is another important energy and information carrier in nature. Without any doubt, exploration of the phonon coherence and its impact on thermal conduction will markedly change many aspects in broad applications for heat control and management in the real world. So far, the application of coherent effect on manipulation of phonon transport is a challenging work. In this article, we review recent advances in the study of the phonon coherent transport in nanomaterials and nanostructures. We first briefly look back the classical and quantum theory of coherence. Next, we review the progresses made in the understanding of phonon coherence in superlattice, nanowires and nanomeshes, respectively, and focus on the effect of phonon coherence on thermal conductivity. Finally, we introduce the recent advances in the direct detection of phonon coherence using optical coherence theory.

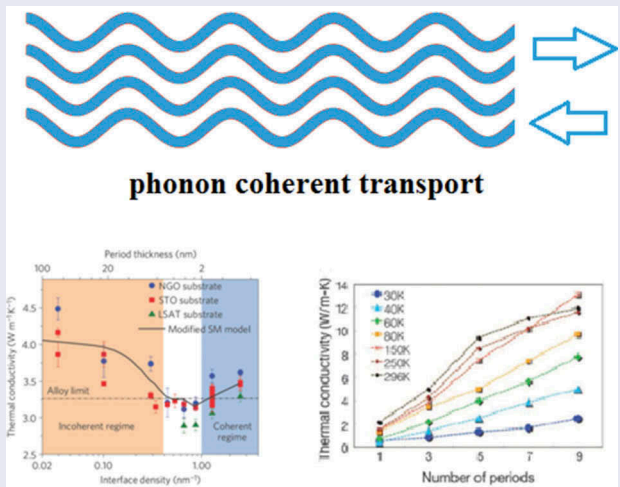

phonon coherent transport

**CONTACT** Gang Zhang ✉ zhangg@ihpc.a-star.edu.sg 🖂 Institute of High Performance Computing, 138632, Singapore

**Abbreviations:** 2D: Two-dimensional; BMS: Brillouin-Mandelstam light scattering; BTE: Boltzmann transport equation; CPT: Coherent population trapping; DFPT: Density functional perturbation theory; DOS: Density of states; EIT: Electromagnetic induce transparency; FFT: Fast Fourier transform; GeNWs: Germanium nanowires; HCACF: Heat current autocorrelation function; HRTEM: High resolution transmission electron microscopy; LJ: Lennard-Jones. MD: Molecular dynamics; MFP: Mean free path; PnCs: Phononic crystals; NMs: Nanomeshs. NWs: Nanowires; SEM: Scanning electron microscopy; SiNTs: Silicon nanotubes; SiNWs: Silicon nanowires; SLs: Superlattices; SMRT: Single mode relaxation time; TEM: Transmission electron microscopy.

## 1. Introduction

The concept of coherence came about when Thomas Young's double-slit optical experiment showed how interference patterns can arise as a result of coherence. With the development of interferometers, coherence has showed up in all wave-like fields from light [1] to acoustics [2]. The advent of quantum mechanics has placed additional responsibility on the double-slit experiment in demonstrating wave-like nature in a variety of physical systems ranging from electrons [3,4] and neutrons [5]. Heat-carrying phonons are high-frequency lattice vibrations in materials ranging from few hundreds gigahertz to tens of terahertz. While there have been experiments demonstrating phonon amplification [6–9] and phonon generation at low temperatures [10], it is only recently that the role of phonon coherence in heat transport has been discussed.

In the past two decades, the tendency of nanoscale devices is increasingly integrated, miniaturized and functionalized, which makes heat conduction in nanostructures becomes one of the most important research focuses due to the potential applications in many areas. Different length scales are associated with phonons and electrons, thereby nanotechnology can significantly reduce the thermal conductivity of the material [11–22], leaving its electrical conductivity almost unaffected [23,24]. Therefore, nanostructures are regarded as potential building blocks for future thermoelectric devices [25–31]. On the other hand, fundamental understanding of heat conduction and nanoscale interfacial thermal resistance in materials, heterojunctions and the interconnects between components is necessary to prevent premature degradation and failure of micro- and nano-electronic devices [32–38].

Different from bulk materials, new phonon physics and novel thermal transport properties arise in low-dimensional nanostructures, like the size dependence of thermal conductivity [39–42] and the confinement of phonons [43–47]. Nanostructured materials, for example, superlattices (SLs), nanowires (NWs) and nanomeshs (NMs), have demonstrated strongly reduced thermal conductivities compared to their parent bulk materials. Because of the high surface-to-volume ratio of nanostructures, the diffusive phonon–boundary scatterings dramatically suppress the mean-free-path (MFP) of phonons. However, it is not uniquely responsible for the remarkable reduction of thermal conductivity in nanostructures. Generally, there are two contrast pictures to describe phonon transport in nanostructures: one is the particle-like incoherent phonons and the other one is the wave-like coherent phonons. When the phonons lose their phase after successive scattering events such as Umklapp phonon–phonon scattering, phonon–impurity scattering and diffusive phonon–boundary scattering, these processes are often referred to as 'incoherent' scattering mechanisms, in contrast to 'coherent' scattering in which the phase is preserved and wave interference is occurred. Most of the conventional approaches to reduce thermal conductivity, such as introduction of rough surface [12,48–52] and impurity scatterings [53–57], are based mainly on incoherent mechanisms by shortening the MFP of phonons. The shortcoming of these approaches is the deterioration of electronic transport properties [12], which limits the dramatic enhancement of thermoelectric performance. Because of the wave nature of phonons, it is possible to control the heat conduction in nanostructures by the interference of phonons. The coherent transport of phonons allows for the control of thermal conductivities in nanostructures with a secondary periodicity such as SLs and NMs by modifying the phonon dispersion relations, thus changing the group velocities and the density of states (DOS) of phonons. In addition, in analogy with electrons and photons, the phononic bandgap, which constitutes the frequency range for which the wave is forbidden to propagate, is formed by interference effects [58]. Thereby, wave interference effects can create fundamentally new approaches for manipulating heat flow.

However, the application of coherent effect on manipulation of phonon transport is a challenging work. On the one hand, for coherent phonon modes to affect thermal transport, the feature size of secondary periodicity should be on a length scale comparable to wavelengths of the phonons that contribute to thermal transport. On the other hand, in order to maintain the phase of coherent phonons, phonons must scatter specularly at the surface boundaries or interfaces of nanostructures. Therefore, nanoscale periodicity and atomically smooth surfaces are required to realize the coherence of phonons at room temperature. With the development of nanotechnology, nanostructures meeting these requirements have been fabricated, such as SLs and NMs, and the experimental observations of coherent heat conduction in these structures have been reported [59–62]. Despite these experimental observations, the range of occurrence of coherent phonon interference is still a controversial question. The issue of coherent versus incoherent transport is important for understanding heat conduction in nanostructures, and recently, the influence of coherent and incoherent scattering mechanisms on thermal conductivity is one of the hottest topics in the field of nanoscale heat transport [61–67].

In this article, we would like to give a review on the present understanding of coherent phonon transport in nanostructures, from both the experimental and the theoretical points of view. The rest of this article is organized as follows. In Section 2, we will briefly outline the classical and quantum theory of coherence. Then, we will discuss optical theory of classical and quantum coherence in heat transport. From Sections 3 to 5, we introduce the recent advances in the study of phonon coherence in SLs, NWs and NMs, respectively, and focus on the effect of phonon coherence on thermal conductivity. We also discuss the conditions under which coherent thermal transport can play dominated role in thermal conductivity of these structures. In Section 6, we introduce the implications of understanding coherence in thermal transport using optical theory of coherence. In Section 7, we present the conclusions and brief outlook.

## 2. Classical and quantum description of coherence

When Thomas Young placed two pinholes in front of a monochromatic source, it was found the superposition of two waves with $I_1$ and $I_2$ at the same point resulted in generating minima and maxima created by the interference of the two waves $\langle I_1 I_2 \rangle$. Subsequently, the coherence of two intensity fields at different points in space and time $I(r_1, t)$ and $I(r_2, t + \tau)$ can be described in a coherence function as

$$g_2(r_1, r_2, \tau) = \frac{I(r_1,t)I(r_2,t+\tau)}{I(r_1,t)I(r_2,t+\tau)} \tag{1}$$

where $\langle \rangle$ implies time averaging. This description relates to the concept of visibility of all optical interferometers [68].

Quantum mechanics came into picture in the early 1900s, and coherence soon became associated with concepts such as particle–wave duality and indistinguishability. However, there exists an ongoing debate on the relation of quantum mechanics to the intuitions of our physical world. The main issue lies in decoherence events that couples to the environment and 'collapses' our quantum coherence between states into localized ones from which classical behavior emerges [69]. Thus, coherence in quantum mechanics is hard to realize experimentally. Nevertheless, the field of quantum optics has developed the concept of optical quantum coherence [70] just like classical coherence in Equation (1), except that intensities are replaced by field operators such that

$$g_2(r_1, r_2, \tau) = \frac{\hat{a}^\dagger(r_1,t)\hat{a}^\dagger(r_2,t+\tau)\hat{a}(r_2,t+\tau)\hat{a}(r_1,t)}{\hat{a}^\dagger(r_1,t)\hat{a}(r_1,t)\hat{a}^\dagger(r_2,t+\tau)\hat{a}(r_2,t+\tau)} \qquad (2)$$

where $\hat{a}^\dagger$ and $\hat{a}$ are the creation and annihilation operators of the optical field, $I = \hat{a}^\dagger\hat{a}$ . For classical coherence of electromagnetic fields, $g_2(r_1, r_2, \tau) = 1$ for perfect coherence and $g_2(r_1, r_2, \tau) = 2$ for incoherent field. In quantum optics, it is possible that the coherence function $g_2(r_1, r_2, \tau) < 1$ in Equation (2) depending on the state of the electromagnetic field and a variety of such states has been generated experimentally under well-controlled experimental conditions [71–74].

There have been various works on generating and detecting coherent heat-carrying phonons in materials [75–78]. Early low-temperature experiments in defect-doped materials have allowed direct observation of spectral diffusion of phonons [79,80] and achieved stimulated emission of phonons [81–83]. The concept of coherent heat-carrying phonons in the quantum regime has also been discussed in quantum phonon optics [84,85] and quantum information [86–88]. However, the effect of coherent phonons on heat transport has only been discussed recently. Obviously, in addition to the importance in fundamental research, the effect of phonon coherence on thermal conductivity of nanomaterials and nanostructures can provide direct features in practical application. So in the next three sections, we will focus on phonon coherence and its effect on thermal properties of various nanomaterials and nanostructures. In Section 6, we will discuss concepts to characterize coherent heat-carrying phonons in terms of coherence functions in Equations (1) and (2).

## 3. Coherent phonon transport in SLs

A phonon SL corresponds to a periodic arrangement of different crystalline materials. Due to the applications in thermoelectric devices and optoelectronic devices, thermal conductivity in semiconductor SLs has been widely investigated numerically [89–93] and experimentally [94–98]. In the picture of incoherent phonon transport, phonons are treated as particles if the phonon MFP is shorter than the period thickness of SLs. Phonons in different layers of an SL are not coherently correlated, thereby each layer is subject to its bulk dispersion relations, and the interface boundary resistance [99] is used as the important feature of an SL. The effective cross-plane thermal conductivity of the SL is then given as [93],

$$\kappa_{SL} = \frac{2L}{L(1/\kappa_1 + 1/\kappa_2) + 2R_B} \qquad (3)$$

where ($L_i$, $\kappa_i$) are the thickness and thermal conductivity of the individual layers and $R_B$ is the thermal boundary resistance. Herein, $L_1 = L_2 \equiv L$ is simply assumed. According to this classical prediction, the thermal conductivity of SLs decreases as the layer thickness L decreases because the phonon MFP is limited to the layer thickness. This trend has been observed in many experimental investigations on different SLs such as those made of Si/Ge, GaAs/AlAs and $Bi_2Te_3$/$Sb_2Te_3$ [95,98–103]. Nonetheless, with a further decrease in layer thickness which is shorter than the phonon MFP, the reverse trend is observed, i.e. the thermal conductivities of shorter period length SLs increase as the layer thickness decreases [95,104,105]. This phenomenon must be explained by the wave nature of phonons. Due to the interference of the phonon waves transporting towards and away from the interfaces, the bulk phonon dispersion relations can be modified. Although effects such as folding, bandgap creation and band flattening reduce the phonon group velocities, they are reduced as the period decreases [93], thereby the thermal conductivity increases as the layer thickness of SLs decreases in the coherent regime.

Applying a lattice dynamics model, Simkin and Mahan showed that the thermal conductivity of SLs had a minimum value for a layer thickness somewhat smaller than the MFP of the phonons, which was the crossover between the particle and wave interference regimes [93]. Using molecular dynamics simulations, Chen *et al.* found that a minimum thermal conductivity would occur if the phonon MFP was comparable to or longer than the period length and the lattice constants of the alternating layers were very close to each other [106]. Using relaxation times that included both anharmonic and interface roughness effects, Garg and Chen computed the thermal conductivity of Si/Ge SLs from density-functional perturbation theory and concluded that for short-period SLs, when SL periodicity increases, interplay between decrease in group velocity and increase in phonon lifetimes will lead to a minimum in the cross-plane thermal conductivity [107].

Due to the decoherence of phonons by defects, interfaces, surface imperfections and anharmonicity, the unambiguous observation of wave behavior is extremely challenging. Some experimental measurements [100–102,108,109] show that heat transport and the presence of a minimum thermal conductivity in SLs strongly depend on the quality of the interfaces. Diffusive scattering dominates heat transport when interfaces are rough, so the minimum point will not be reached. Recently, a new approach was proposed by Ravichandran *et al.* by using high-quality oxide SLs grown through molecular beam epitaxy [60], as shown in Figure 1(a). Ravichandran *et al.* observed a minimal thermal conductivity as a function of periodicity, provided evidence for the existence of wave interference and phononic bandgap effects on heat transport and presented an unambiguous demonstration of the theoretically predicted crossover from particle-like to wave-like phonon scattering in epitaxial perovskite oxides SLs [60]. The plots in Figure 1(b) and (c) can be clearly divided into two regimes based on whether the thermal conductivity increases (coherent) or decreases (incoherent) with increasing interface density (reciprocal of periodicity). In the low interface density regime, the system can be modeled as a series of bulk thermal resistances with the resistance of the interfaces added in series with the bulk resistances. In this incoherent regime, the behavior of the phonons is particle-like, and hence, the thermal resistance increases linearly with increasing interface density. At the high interface density regime, where the SL period is comparable to the coherence length of the phonons, the wave nature of phonons must be considered. In this limit, the phonon dispersion of SL forms mini-bands owing to band folding along the cross-plane axis, which decreases the overall

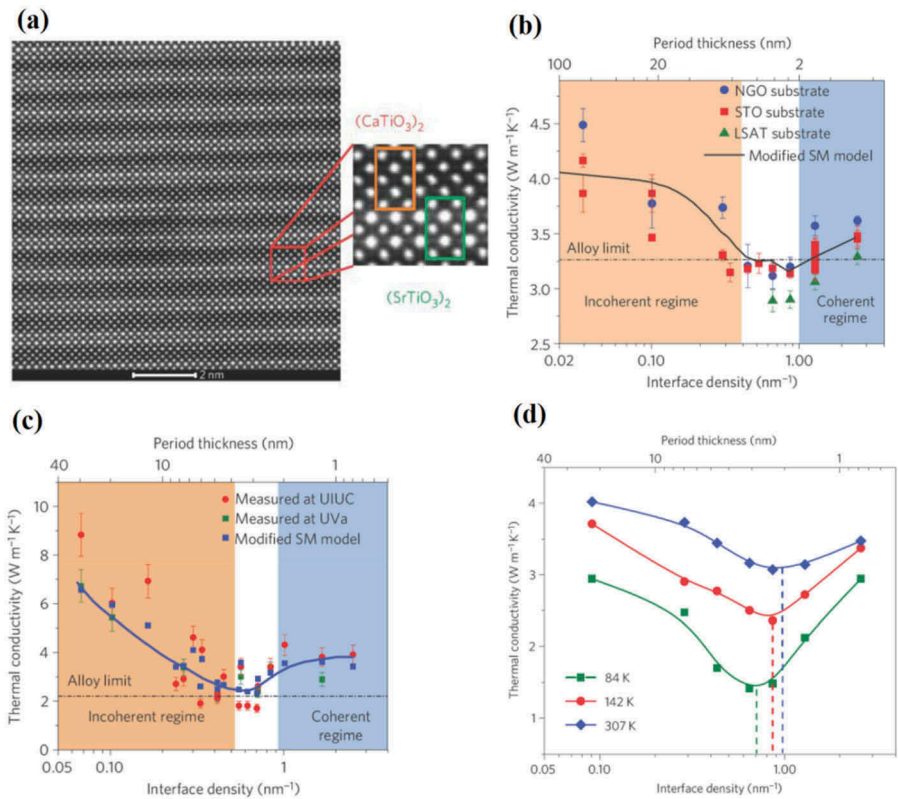

**Figure 1. Crossover from incoherent to coherent phonon transport in perovskite oxides SLs.** (a) Transmission electron microscopy image of an $SrTiO_3/CaTiO_3$ superlattice with atomically sharp interfaces. (b) $(STO)_m = (CTO)_n$ SL, (c) $(STO)_m = (BTO)_n$ SL, measured thermal conductivity values for SLs as a function of interface density at room temperature. (d) Measured thermal conductivity values for $(STO)_m = (CTO)_n$ SLs as a function of interface density at different temperatures.

Note: Adapted from Ref. [60].

group velocity of phonons. As the interface density increases, the number of mini-bands decreases, and this leads to an increase in average phonon group velocity; therefore, the thermal conductivity increases with increasing interface density. Figure 1(d) shows a deeper and clearer minimum for the thermal conductivity when the temperature is reduced from 307 to 84 K, which provides evidence for stronger interference effects at lower temperatures.

From Equation (3), it is indicated that if the phonons are scattered diffusely at the interfaces of SL, the constituent layers have their own defined thermal conductivities that can be added in series to obtain the cross-plane thermal conductivity of the SL, which will not depend on the number of layers. However, if the phonons preserve their phases when crossing the interfaces, interference effects may therefore develop and the thermal conductivity will be linearly proportional to the total thickness of SL. Luckyanova *et al.* measured the thermal transport properties of SLs with a constant periodicity but a varying number of periods [97]. They fabricated five GaAs/AlAs SLs using metal–organic chemical vapor deposition, with periods of 1, 3, 5, 7 and 9, where each period consisted of a 12-nm GaAs and a 12-nm AlAs layer as shown in Figure 2(a). Measured thermal conductivity by time-domain thermal reflectance as a function of the number of periods at temperatures from 30 to 300 K is demonstrated in Figure 2(b). The thermal conductivity has linear dependence on total SL thickness from 30 to 150 K. This nearly linear dependence of thermal conductivity on number of SL periods presents the clear evidence of coherent phonon transport through the layers, whereas the nonlinear dependence at temperatures greater than 150 K suggests the increased influence of incoherent effects. However, because the thermal conductivity

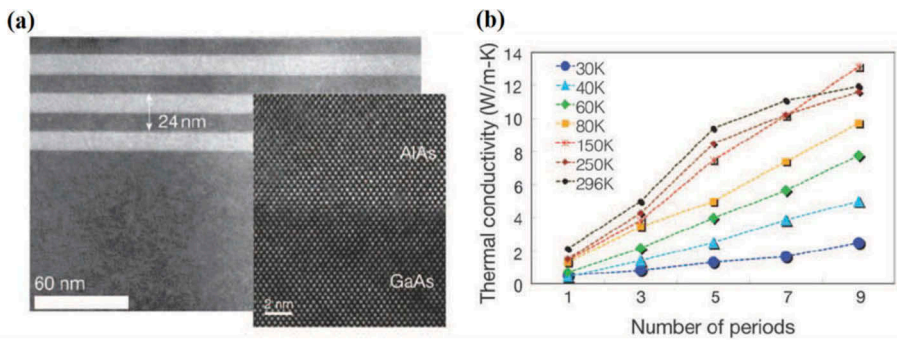

**Figure 2. Wave-like heat transport in GaAs/AlAs SLs.** (a) Cross-sectional transmission electron microscope (TEM) image of the three-period SL. (Inset) High-resolution transmission electron microscopy (HRTEM) image of one of the interfaces. (b)Measured thermal conductivity of GaAs/AlAs SLs as a function of number of periods.

Note: Adapted from Ref. [97].

still increases with the period number even at 296 K, coherent phonons in SLs still conduct a considerable fraction of heat even at room temperature.

To understand the experimental results, Luckyanova *et al.* used the first-principles-based lattice dynamics approach to extract detailed information such as the phonon MFP distribution and the MFP dependence of the thermal conductivity [97]. In their first-principles calculation, the thermal conductivity of GaAs/AlAs SLs was computed by solving the Boltzmann transport equation in the single-mode relaxation time (SMRT) approximation. The thermal conductivity was computed by summing over the heat conducted by all the phonon modes ($\lambda$) in the Brillouin zone, given by Equation (4),

$$k_\alpha = \frac{h^2}{N\Omega k_B T^2} \sum_\lambda c_{\alpha\lambda}^2 \omega_\lambda^2 \overline{n_\lambda}(\overline{n_\lambda} + 1)\tau_\lambda \tag{4}$$

The scattering rate, $1/\tau_\lambda$, of a phonon mode $\lambda$ is taken to be the sum of a term describing scattering due to interfacial disorder ($1/\tau_{\lambda a}$) and a term describing anharmonic scattering ($1/\tau_{\lambda b}$) as in Matthiessen's rule. The anharmonic scattering rates ($1/\tau_{\lambda b}$) are computed using the lowest-order three-phonon scattering processes in the SMRT approximation via

$$\begin{aligned} \frac{1}{\tau_{\lambda b}} = \pi \sum_{\lambda'\lambda''} &\left| V_3(-\lambda, \lambda'\lambda'') \right|^2 \times \\ &[2(\bar{n}_{\lambda'} - \bar{n}_{\lambda''})\delta(\omega(\lambda) + \omega(\lambda') - \omega(\lambda'')) + \\ &(1 + \bar{n}_{\lambda'} + \bar{n}_{\lambda''})\delta(\omega(\lambda) - \omega(\lambda') - \omega(\lambda''))] \end{aligned} \tag{5}$$

where $V_3(-\lambda, \lambda'\lambda'')$ is the three-phonon coupling matrix elements or the weighted Fourier transforms of the cubic force constants. Interface roughness is simulated as a random mixing of Ga and Al atoms in a narrow region around the interface. According to the perturbation theory, the scattering rates due to interfacial disorder are calculated using:

$$\frac{1}{\tau_{\lambda a}} = \frac{\pi}{2N} \omega_\lambda^2 \sum_{\lambda'} \delta(\omega_\lambda - \omega_{\lambda'}) \sum_\sigma g(\sigma) \left| re(\sigma|\lambda') re(\sigma|\lambda) \right|^2 \tag{6}$$

All ingredients necessary to compute the thermal conductivity, including the phonon frequencies, group velocities, populations and lifetimes, are obtained from first-principles calculation using density functional perturbation theory (DFPT). The second-order and third-order interatomic force constants used to estimate the above parameters are also obtained from DFPT. Supplementary materials for Ref. [97] provide more computational details.

The anharmonic and interface scattering rates at different frequencies are showed in Figure 3(a). The interfacial scattering of high-frequency

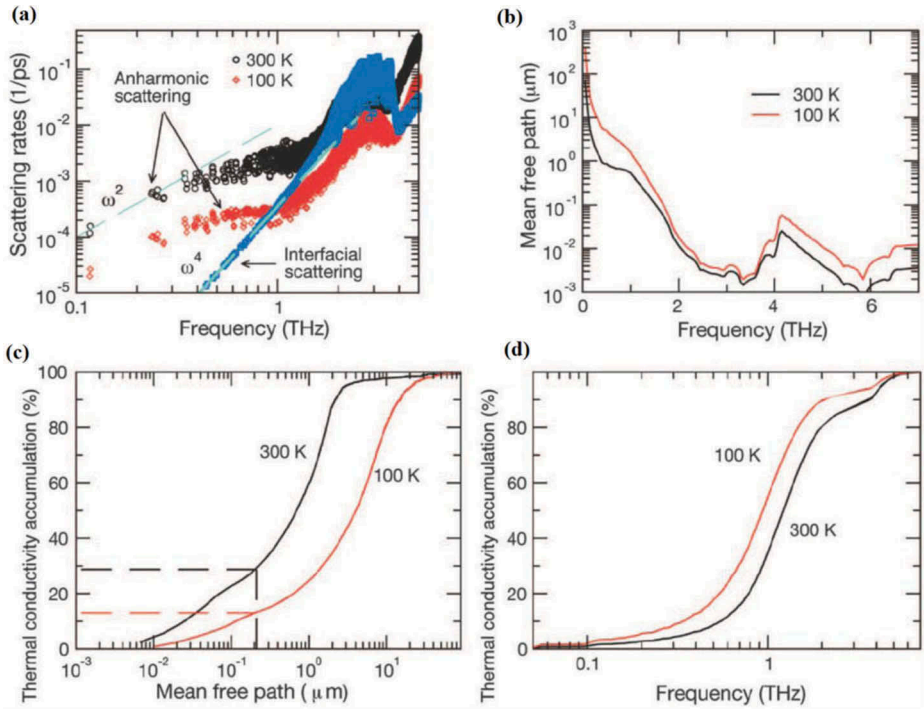

**Figure 3.** First-principles calculation results for an infinite 12 nm by 12 nm GaAs/AlAs SL. (a) Comparison between anharmonic (red for 100 K and black for 300 K) and interface (blue) scattering rates. Dashed blue lines are the fits to the scattering rates describing the $\omega^2$ and $\omega^4$ behaviors of the anharmonic and interfacial scattering, respectively, in the low-frequency regime. (b) Phonon MFPs in the SL as a function of frequency. Thermal conductivity accumulation in the SL as a function of (c) phonon MFP and (d) frequency, at 100 and 300 K.

Note: Adapted from Ref. [97].

phonons lead to a reduction in their heat-carrying ability and caused an overall decrease in the thermal conductivity of SLs. Low-frequency phonons have long MFPs as shown in Figure 3(b) and can ballistically propagate through the entire SL structure whose thickness was shorter than the MFPs. The contribution of these long MFP phonons to the total thermal conductivity is shown in Figure 3(c) and (d), where thermal conductivity accumulation is plotted against MFP and frequency, respectively. Phonons with MFPs longer than 216 nm, the thickest measured SL, contributed 87% (at 100 K) and 71% (at 300 K) to the total thermal conductivity of an infinite SL, as indicated by the dashed lines in Figure 3(c). Therefore, the long-MFP coherent phonons play dominant role in heat transport through these SLs.

Very recently, nanotwinned structure has attracted a plenty of research interest in thermal transport community [110–116]. Zhou *et al.* found a new strategy for enhancing the figure-of-merit of thermoelectrics by decoupling the electronic and phononic transport in nanotwinned structure [110]. Yu *et al.* achieved simultaneous optimization of electrical and thermal transport properties of $Bi_{0.5}Sb_{1.5}Te_3$ thermoelectric alloy by twin boundary engineering [111]. The phonon scattering at twin boundaries has been investigated experimentally [112]. It was found that due to the coherent characteristic of twin boundary, phonon scattering at twin boundary is noticeably weaker with respect to the grain boundary [112]. Dong *et al.* performed molecular dynamics simulations to investigate thermal conductivity of twinned diamond [113]. They found that twin boundaries can indeed lead to additional twin boundary thermal resistance, but the weak phonon scattering only causes a slight reduction in thermal conductivity [113].

## 4. Coherent phonon transport in NWs

Semiconductor NWs have attracted considerable attention due to their potential applications in many areas, such as electronic devices [117,118], biosensors [119,120], thermoelectric devices [121–124] and optoelectronic devices [125–128]. The interest of investigating phonon transport in SiNWs has been greatly stimulated [129–145] since experiments [12,29] revealed that an approximately 100-fold reduction in thermal conductivity over bulk Si can be achieved in SiNWs, while the electrical conductivity and electron contribution to Seebeck coefficient are still similar to those of bulk silicon, which indicated that SiNWs could be applied as high-performance nanoscale thermoelectric materials. Further reducing thermal conductivity of SiNWs is critically important for achieving higher thermoelectric performance. Besides the approaches based on incoherent mechanisms to reduce the thermal conductivity of SiNWs, such as introduction of rough surface [57,145] and defect scatterings [130,142], using the wave nature of thermal phonons is also an effective method to tune the thermal conductivity of SiNWs, such as phonon coherent resonance and phonon confinement.

By equilibrium molecular dynamics simulations, Chen *et al.* observed phonon coherent resonance in Ge/Si core–shell NWs [146]. The coherence of phonons can be probed by the heat current autocorrelation function (HCACF). As shown in Figure 4(a), for both SiNWs and SiNTs, there is a very rapid decay of HCACF at the beginning, followed by a long-time tail with a much slower decay. However, a nontrivial oscillation up to a long time appears in HCACF for Ge/Si core–shell NWs. The long-time region of HCACF reveals that this nontrivial oscillation is not random but shows a periodic manner, showed in inset of Figure 4(a). To understand the mechanism, Figure 4(b) shows the fast Fourier transform (FFT) of normalized HCACF for core–shell NWs, which are very similar to the spectrum of

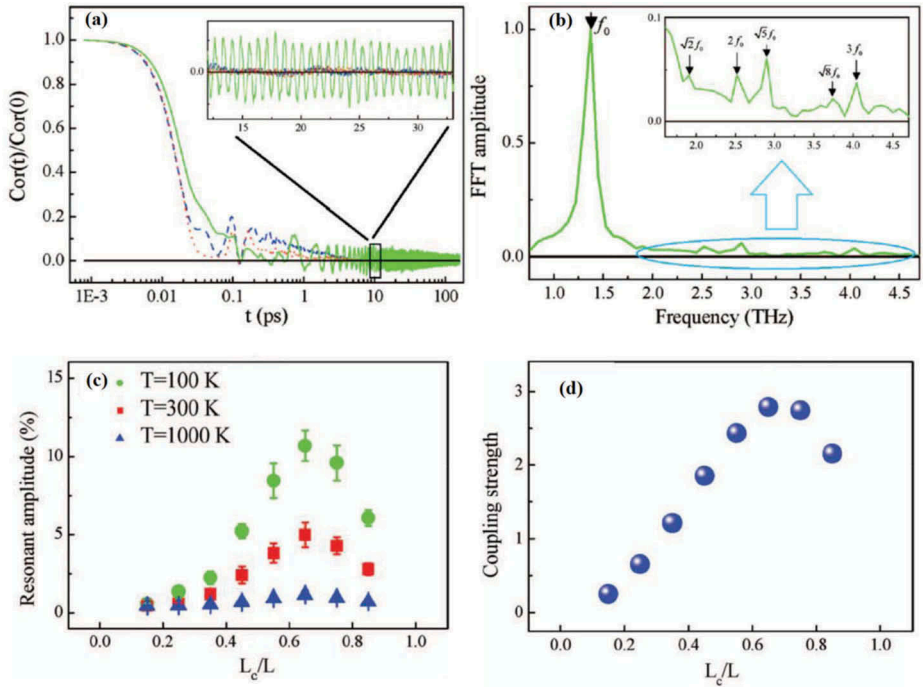

**Figure 4. Phonon coherent resonance in Ge/Si core–shell NWs.** (a) Time dependence of normalized HCACF for SiNWs (blue dashed line), SiNTs (red dotted line) and Ge/Si core–shell NWs (green solid line). Inset shows the long-time region. (b) Amplitude of the FFT of normalized HCACF for Ge/Si core–shell NW in the long-time region. Inset shows the high-frequency oscillation peaks. (c) Oscillation amplitude versus core–shell ratio $L_c/L$. (d) Coupling strength versus core–shell ratio in Ge/Si core–shell NWs.

Note: Adapted from Ref. [146].

the coherent resonance effect in a confined structure. The FFT analysis of HCACF suggests that the nontrivial oscillation is caused by the phonon coherent resonance effect in the transverse direction. In core–shell NWs, atoms on the same cross-section plane have a different sound velocity in the longitudinal direction. As a result, there is a strong coupling of vibrational modes between the longitudinal and transverse vibrations. When the frequency of the longitudinal phonon mode is close to the eigenfrequency of the transverse mode, coherent resonance occurs. Thus, the coherent resonance effect in the transverse direction can indeed manifest itself in HCACF along the longitudinal direction in Ge/Si core–shell NWs. However, in single-component SiNWs and SiNTs, atoms on the same cross-section plane have the same sound velocity, and the transverse motion is decoupled from the longitudinal motion, so that the coherent resonance effect is absent. Moreover, as the transverse phonons are non-propagating, this resonance effect can significantly hinder the heat transport in the longitudinal direction. The phonon coherent resonance effect in

core–shell NWs is structure- and temperature-dependent, as shown in Figure 4(c). When the core size increases, the resonance effect becomes stronger, reaches its maximum amplitude and then decreases. For a given core–shell structure, the oscillation amplitude becomes larger at a lower temperature, as the resonance effect of acoustic wave is a coherent process that requires long-time correlation and the stronger anharmonic phonon–phonon scattering at high temperature causes phonon to lose coherence and leads to the vanishing of the oscillation effect at high temperature. The coupling strength is defined to quantitatively characterize the coupling between the transverse and longitudinal modes in the core–shell structure. As shown in Figure 4(d), the dependence of coupling strength on core–shell ratio agrees qualitatively well with the variation of the measured oscillation amplitude as shown in Figure 4(c). This good agreement reveals that the structure dependence of the oscillation amplitude is caused by the structure-dependent coupling strength.

It is worth emphasizing that in molecular dynamics (MD) simulation, all the modes are equally exited. This is valid when the system temperature is higher than the Debye temperature because the associated energy for each mode can be approximated to the classical case, which means all the modes contribute equally to the total energy. However, at low temperature (with respect to Debye temperature), the freezing out of high-frequency modes in the quantum system makes energy associated with different modes deviates from the classical case. In this case, quantum correction is necessary to qualitatively account for this discrepancy. Several methodologies have been proposed by equating the total energy in the classical and quantum descriptions [147]. For example, a system-level mapping between the classical temperatures used in MD simulation ($T_{MD}$) and the quantum temperature for the real environment ($T_{real}$). This phenomenological quantum correction technique can qualitatively capture the temperature effect on thermal transport in crystalline solids. However, since this method is done based on the system level, it cannot reproduce the actual microscopic properties at low temperature, for example, relaxation time and occupation number, which are govern by the laws of quantum mechanics [148].

The coherent resonance effect and mode coupling provide a new avenue to reduce the thermal conductivity of low thermal conductivity materials even by coating with high thermal conductivity materials, which has been demonstrated by results based on nonequilibrium molecular dynamics simulations of Chen *et al.* [149]. The coating configuration of GeNWs is shown in Figure 5(a). Figure 5(b) demonstrates that when the coating thickness is less than certain critical value, thermal conductivity of Ge/Si core–shell NWs is smaller than that of pristine GeNWs. Figure 5(c) and (d) compares the participation ratio for eigenmodes in pristine GeNWs,

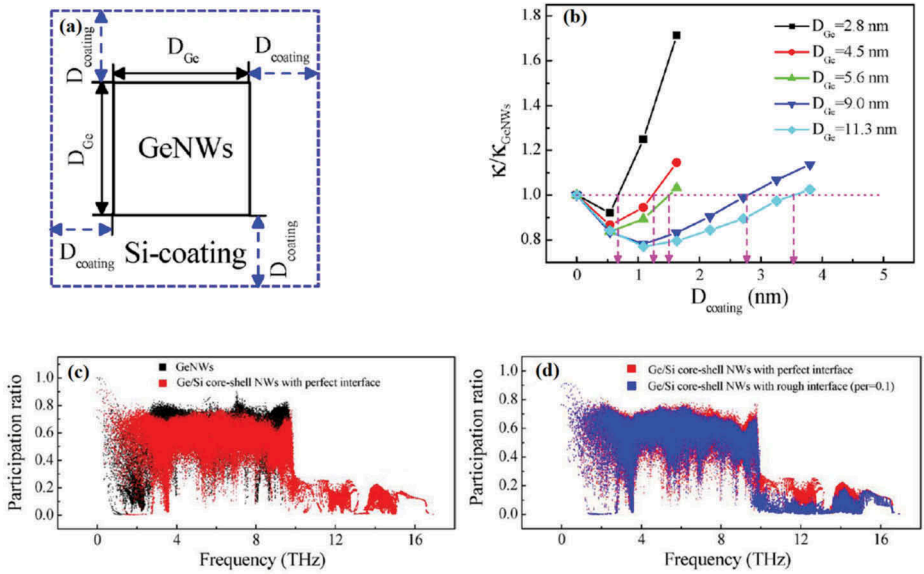

**Figure 5.** Thermal conductivity of Ge/Si core–shell NWs. (a) Coating configuration of GeNWs. (b) Normalized thermal conductivity versus coating thickness for different $D_{Ge}$. Thermal conductivity of GeNWs at each $D_{Ge}$ is used as reference. The dashed arrows point the critical coating thickness when thermal conductivity of Ge/Si core–shell NWs ($\kappa_{core-shell}$) is equal to that of GeNWs. The dashed line is drawn to guide the eye. (c) and (d) P-ratio for different phonon modes in GeNWs before and after coating. The black, red and blue denote, respectively, participation ratio in GeNWs, Ge/Si core–shell NWs with perfect interface and Ge/Si core–shell NWs with 10% interfacial roughness.

Note: Adapted from Ref. [149].

Ge/Si core–shell NWs with perfect interface and Ge/Si core–shell NWs with rough interface. The coherent resonance induced by coupling between the transverse and longitudinal modes in core–shell NWs reduces the participation ratio of low-frequency longitudinal acoustic phonons. As the strongest resonance peak is related to the eigenmode with lowest frequency in transverse direction, the localization is remarkable for phonons with long wavelength (>NW diameter). Phonon scattering by the rough interface induces an additional reduction of participation ratio in the high-frequency regime (>10 THz). The interface roughness results in the additional localization effect of optical phonons, thus leading to the further reduction of thermal conductivity compared to the perfect interface.

Recently, Wingert *et al.* developed an experimental technique with drastically improved sensitivity capable of measuring thermal conductance values down to ~10 pW/K [43]. Based on this platform, they found that at room temperature the thermal conductivities of Ge/Si core–shell NWs were lower than those of GeNWs with the same size [43] and explained this phenomenon by the nontrivial phonon coherent resonance effect mentioned above. Wingert *et al.* also found that thermal conductivities of the sub-20-nm diameter GeNWs were remarkably lower than the calculated results by Boltzmann transport models using bulk dispersions, as shown in Figure 6(a), and ascribed the ultra-low thermal conductivity of thin GeNWs to the phonon confinement effect [43]. From the characteristic lengths, phonon MFP and wavelength, phonon transport can be categorized into three regimes: bulk-like, boundary scattering (or Casimir) and confinement [46]. For Si and Ge nanostructures, the size region of sub-30 nm is considered as the 'confinement regime'. Due to spatial confinement of phonons, the acoustic dispersions are modified, and the group velocities are significantly lower than those of bulk counterpart as shown in Figure 6(b), which lead to the remarkable reduction of thermal conductivity. Very recently, Kargar *et al.* directly observed confined acoustic phonon polarization branches in free-standing GaAs NWs [150]. Based on excellent agreement of the dispersions obtained from Brillouin–Mandelstam light scattering experiments and calculated for the exact NW shape and material parameters, Kargar *et al.* proved the confined nature of the phonons [150].

Recent experiments [151–155] show that twin planes are commonly formed in NWs, including InP, SiC, GaP and Si NWs. These coherent twin boundaries form periodic lamellar twinning along the NWs and form a twinning SL NW. The diameter and period length of these metamaterial NWs can be controlled during the synthesis process, offer-

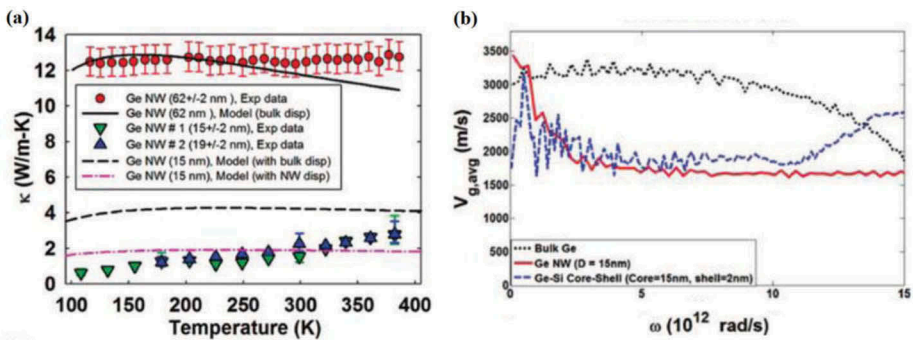

**Figure 6.** Thermal conductivity of GeNWs, comparisons between modeling and experimental results. (a) The calculated κ of the 62 nm GeNW using bulk dispersions (black solid line) agrees well with the experimental data (circles), whereas the calculated κ of the 15 nm GeNW using bulk dispersions (black dashed line) is significantly higher than the experimental data (triangles), by more than 100% below 250 K. Modeling results using NW dispersions (pink dash-dot line) show good agreement with the data. This suggests the important role of phonon confinement in small diameter NWs. (b) Calculated group velocities of 15 nm GeNWs (red solid line) and a 19-nm Ge–Si core–shell NWs (15 nm Ge core + 2 nm thick Si shell, blue dashed line) are significantly lower than that of bulk Ge (black dotted line).

Note: Adapted from Ref. [43].

ing the degree of freedom for tuning their properties. Such unique characteristics are of practical interest for high-performance thermo-electrics. The impact of twinning on heat conduction in NWs has attracted a plenty of research interest [156–159]. Using atomistic simu-lations, Porter *et al.* reported a crossover from diffusive interface scat-tering to SL-liked behavior for phonon transport across twin boundaries in SiNW [156]. By nonequilibrium molecular dynamics simulations, Xiong *et al.* found that the thermal conductivity of the Si metamaterial NWs decreases firstly with twinning SL period increases, reaching a minimum value and then progressively increases [157]. This is similar to that observed in the heterostructure SLs. The minimum thermal conductivity in the heterostructure SLs is a consequence of the interplay between the phonon coherence and the interface scattering, which is discussed in Section 3 of this review. However, because of the weak phonon scattering effect at twin boundaries, the underlying mechanism in the twinning SL NWs completely differs from the one observed in heterostructure SL. Based on phonon mode analysis, it is suggested that the minimal thermal conductivity origins from the disappearance of favored atom polarization directions [157].

## 5. Coherent phonon transport in NMs

In recent years, within nanostructuring strategies, an intriguing concept is the periodically porous structure. Such a structure called as NM is com-posed of a host material from which material is removed in the form of spatially periodic holes to create macroscopic arrays of nanoscale holes in two-dimensions (2D). Manipulating the phonon spectrum with 2D NMs has attracted a great deal of research interest [20,23,39,61–67,160–171]. Reproducible low thermal conductivity, sufficient electrical properties and good mechanical strength make 2D silicon NM as a promising candidate for thermoelectric applications, such as electrical power generation and on-chip thermal management for solid-state devices [160]. Although the impact of nanoscopic holes on reduction of thermal conductivity is widely accepted, the underlying mechanism is still not clear.

The periodic holes in the NM introduce a secondary artificial periodicity to the original lattice, potentially modifying the phonon dispersion rela-tions. From the viewpoint of coherent transport, the resulting phononic bandgaps and reduced group velocities would suppress the thermal con-ductivity. It has been argued that coherent phonons affect thermal trans-port in NMs, and the range of occurrence of such phonon interference is a long-lasting question [61,63–67]. Recently, the influence of coherent and incoherent scattering mechanisms on thermal conductivity of NMs is one of the hottest topics in the field of nanoscale heat transport. Phase preservation is necessary for phonon coherent transport. To maintain phonons' phase, specular reflection by surfaces is needed; therefore, heat-carrying phonon wavelengths should be much larger than the roughness of surfaces. NMs can be fabricated via lithography techniques, but the atomically smooth surfaces are challenging to achieve. At very low temperatures below 1 K, heat-carrying phonon wavelengths are expected to increase by two orders of magnitude with respect to those at room temperature, which weakens the requirement for high-quality interfaces. Zen *et al.* experimentally demonstrated the impact of the coherent effects on thermal conduction reduction in silicon nitride membranes at sub-Kelvin temperature where phonon wavelengths were longer than the characteristic size of the structures [62]. They fabricated two structures with different period $a = 0.97$ μm and a = 2.42 μm but same filling factor of holes and measured the thermal conductance of full membrane and two square phononic crystal (PnC) samples. As shown in Figure 7(b), the thermal conductance of PnC is much lower than that of full membrane. This remarkable reduction of thermal conductance cannot be simply ascribed to the porosity correction factor, which describes the reduction in conductance for bulk porous materials. The possibility that thermal conductance is limited by diffusive scattering inside bulk SiN is excluded by the remarkable difference in the temperature dependence between the full membrane and the two PnC samples as shown in Figure 7(b). In addition, if diffusive

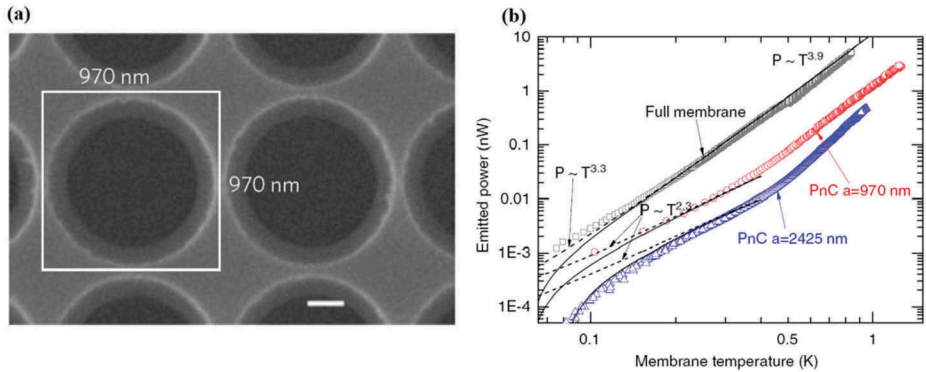

**Figure 7. Thermal conductivity of periodic porous films.** (a) Scanning electron microscopy image of a periodic porous film made from a square array of air cylinders with $a = 970$ nm on a silicon nitride matrix. Scale bar, 200 nm. (b) Measured emitted phonon power versus temperature. Gray squares show the data for the full, uncut membrane; red circles ($a = 0.97$ μm) and blue triangles ($a = 2.42$ μm) show the data for the two square PnC samples. Theoretical calculations with and without back radiation from the substrate are shown by solid and dashed lines, respectively. Agreement between theory and experiments shows that internal and diffusive scattering at interfaces do not play a significant role in heat flow and that wave interference is the main mechanism that affects phonon transport at low temperatures.

Note: Adapted from Ref. [62].

surface scattering at the hole edges dominated, the longer-period PnC sample should produce a higher, instead of a lower, thermal conductance, as the neck dimensions are larger but the edge roughness the same.

From kinetic theory, a simple formula for the thermal conductivity $\kappa$ can be written as:

$$\kappa = \sum_j \int_0^{\omega_{max}} \frac{\hbar^2 \omega^2}{k_B T^2} \frac{e^{\hbar\omega/k_B T}}{(e^{\hbar\omega/k_B T} - 1)^2} g_j(\omega) v_j(\omega) l_j(\omega) d\omega \tag{7}$$

where $\hbar$ is Planck's constant, $k_B$ is the Boltzmann constant, $j$ are different polarizations, $T$ is the temperature, $g_j$ is the DOS, $v_j$ is the group velocity and $l_j$ is the mean free path. The main effect of wave interference is to modify the phonon dispersions which control the propagation of phonons. In addition, in analogy with electrons and photons, interference effects give rise to forbidden energy gaps for phonons. In order to explore the underlying coherent phonon mechanism of thermal conductivity reduction, dispersion relations of full membrane and two periodic holey structures were calculated by Zen et al. [62], as shown in Figure 8(a)–(c). Figure 8(d) and (e) compares the phonon DOS and group velocity, respectively. In Figure 8(b), the PnC membrane with $a = 0.97$ μm has the band gap (width 0.7 GHz) at frequency 3.3 GHz. On the other side, in Figure 8(c), the other PnC sample with $a = 2.42$ μm has negligible band gap. Unintuitively, as shown in Figure 7(b), the larger-period PnC actually has a lower thermal conductance than the smaller-period structure, which has the maximal band gap. Therefore, maximizing the band gap does not necessarily lead to minimum thermal conductance. It is clearly demonstrated in Figure 8(e) that the average group velocities in the PnC membranes are much smaller than those in the full membranes for the whole range of phonon frequencies which contribute to the thermal transport at sub-Kelvin temperature. Given in Equation (7), thermal conductivity is proportional to the product of the phonon DOS and the average group velocity at each energy. Therefore, the main impact on the reduction of thermal conductance of PnCs comes not from the existence of a band gap directly, but from the combination of the reduction of the group velocities and the phonon DOS.

Although Zen et al. found that phonon transport in the PnC is strongly suppressed at sub-Kelvin temperature by the coherent modification of the phonon band structure, their results may still have relevance to room temperature thermal transport [62] because it is reported that a large proportion of heat, even at room temperature, is carried by phonons with relatively long mean free paths >1 μm [172–174]. There is considerable effort to explore the influence of coherence on thermal conductivity of silicon NMs at room temperature [20,23,61–64,175]. Yu et al.

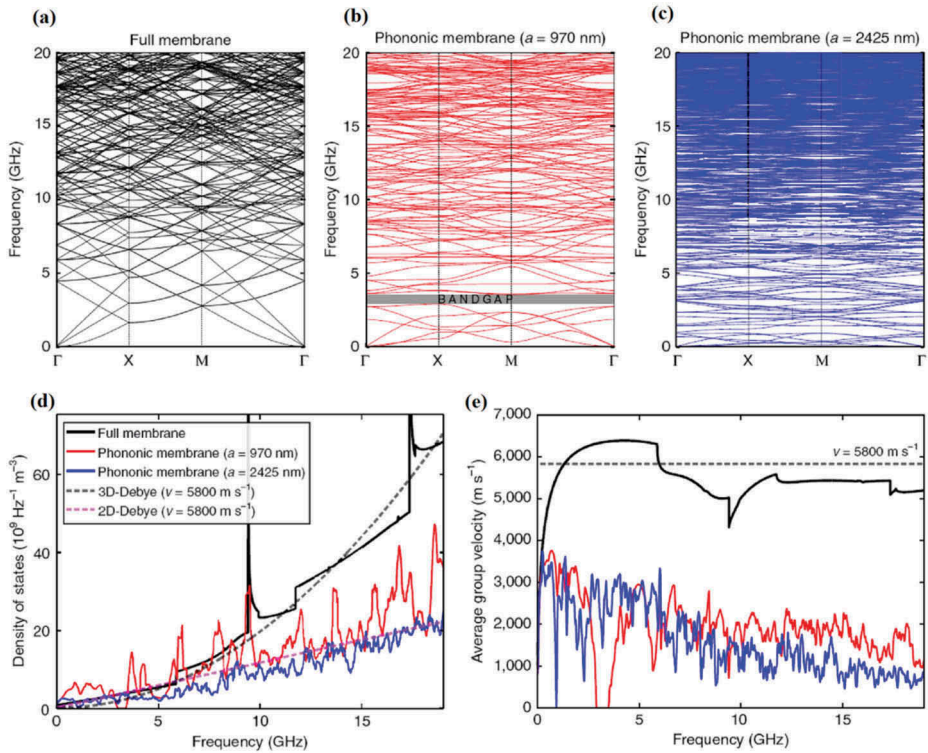

**Figure 8. Band structure DOS and group velocity.** Dispersion relations (band structure) in the main symmetry directions of the first Brillouin zone (BZ) for the SiN (a) full membrane, (b) square lattice PnC with $a$ = 0.97 μm and (c) square lattice PnC with $a$ = 2.42 μm. Complete band gap is observable at 3.3 GHz for the PnC with $a$ = 0.97 μm. (d) The corresponding densities of states (PnC $a$ = 0.97 μm, red, PnC $a$ = 2.42 μm, blue, full membrane, black) with 2D (pink dash) and 3D (gray dash) Debye models. (e) Average group velocity (averaged over the whole 2D BZ) for full (black) and the two PnC membranes (PnC $a$ = 0.97 μm, red, PnC a = 2.42μm, blue). The PnC DOS and group velocity curves have been smoothed for visual clarity.

Note: Adapted from Ref. [62].

experimentally found that the thermal conductivity of silicon NM was consistently reduced by 2 orders of magnitude and approaches the amorphous limit at room temperature [23]. Some groups ascribed the ultra-low thermal conductivity of silicon NMs to coherent phonon interaction [20,23], as analogy with that has been found in the SL structures. Dechaumphai and Chen [162] have reported simulation results of partially coherent transport to show good agreement with the data of Yu *et al.* Alaie *et al.* also proposed a hybrid thermal conductivity model which assumed partially coherent and partially incoherent boundary scattering to reproduce the experimental values of themselves [61]. On the other hand, it has been demonstrated that diffuse phonon scattering dominates heat transport in periodic porous silicon membranes at room temperature due to the short wavelengths of heat-carrying phonons [66,161]. Monte Carlo simulation results of Jain *et al.* provide strong support to the 'necking effect' [176] in incoherent transport as an important factor behind the reduced thermal conductivities for limiting dimensions >100 nm [66]. However, the results for the smaller feature size still remain puzzling. Based on the assumption of a disordered surface, Ravichandran and Minnich fitted the ultra-low thermal conductivity against the data of Yu *et al.* by Monte Carlo simulations [161]. To uncover the impact of incoherent phonon–boundary scattering, the neck size effect was carefully analyzed by Lim *et al.*, and it was showed that the neck size in the range of 16−34 nm impacted phonon transport greatly at 300 K [65].

These various interpretations originate from differing views of the criterion that whether wavelengths or Umklapp scattering-limited mean free paths are the coherence length scales of phonons in periodic structures. It is uncontested that the incoherent models can break down when periodicities $p$ are smaller than the dominant phonon wavelengths $\lambda$, and the coherence effects can be safely neglected when length scales are larger than Umklapp scattering-limited mean free paths $\Lambda_U$. However, there is considerable debate on the role played by phonon coherence when periodicities are large compared with $\lambda$ but small compared with $\Lambda_U$. Since incoherent scattering may reproduce the ultra-low thermal conductivity of NMs due to surface disorder and partially coherent transport may also result in the same without consideration of disorder, the effect of coherence at room temperature may be difficult to resolve without additional data and characterization [177]. Very recently, Lee *et al.* isolated the wave-related coherence effects by comparing periodic and aperiodic silicon NMs and quantified the backscattering effect by comparing variable-pitch NMs [67]. The schematic illustration of periodic and aperiodic silicon NMs is shown in Figure 9(a). They measured identical thermal conductivities for periodic and aperiodic NMs of the same average pitch, as shown in Figure 9(c). Because coherence effects would be disrupted by the aperiodicity, Lee *et al.* concluded that phonon coherence was unimportant for thermal transport in silicon NMs with periodicities of 100 nm and higher and temperatures above 14 K [67]. In addition, a clear $T^3$ trend is observed at low temperature, consistent with the classical diffuse boundary scattering theory. Ray tracing simulations support the measurement results, as shown in Figure 9(c), and the experimental results agree well with the particle model using the full diffusive phonon–boundary scattering. Therefore, phonon backscattering, as manifested in the classical size effect, is responsible for the thermal conductivity reduction [67].

Furthermore, Lee *et al.* deeply investigated the backscattering effect in variable-pitch NMs [67]. Figure 10(a) shows scanning electron microscope (SEM) images of silicon NMs with pitches ($p_x \times p_y$) of 100 × 100 nm, 200 × 100nm and 1000 × 100 nm, with similar neck size and thickness. The

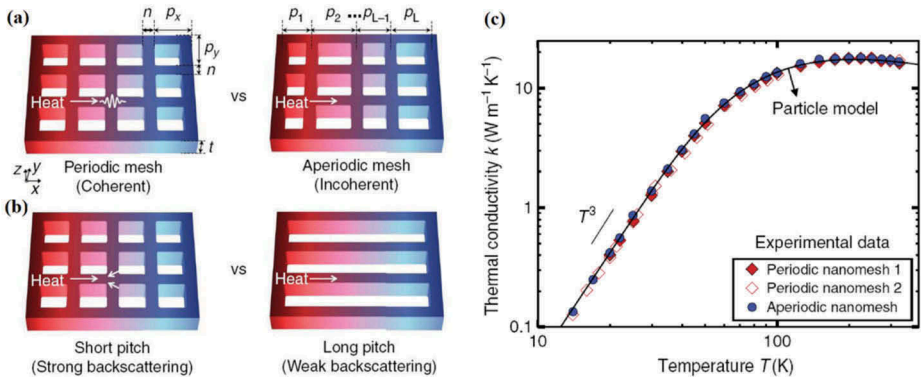

**Figure 9. Isolating coherence effects with periodic and aperiodic NMs.** Schematic illustration of periodic and aperiodic silicon NMs (a); short-pitch and long-pitch NMs (b). (c) Experimental data (points) and the Boltzman transport equation (BTE) particle model with diffuse surfaces (line) show excellent agreement for $\kappa(T)$ of two periodic and one aperiodic NMs. The very similar $\kappa$ between the three samples at all $T$ indicates negligible coherence effects for thermal transport in silicon NMs for $p \geq 100$ nm and $T > 14$ K.

Note: Adapted from Ref. [67].

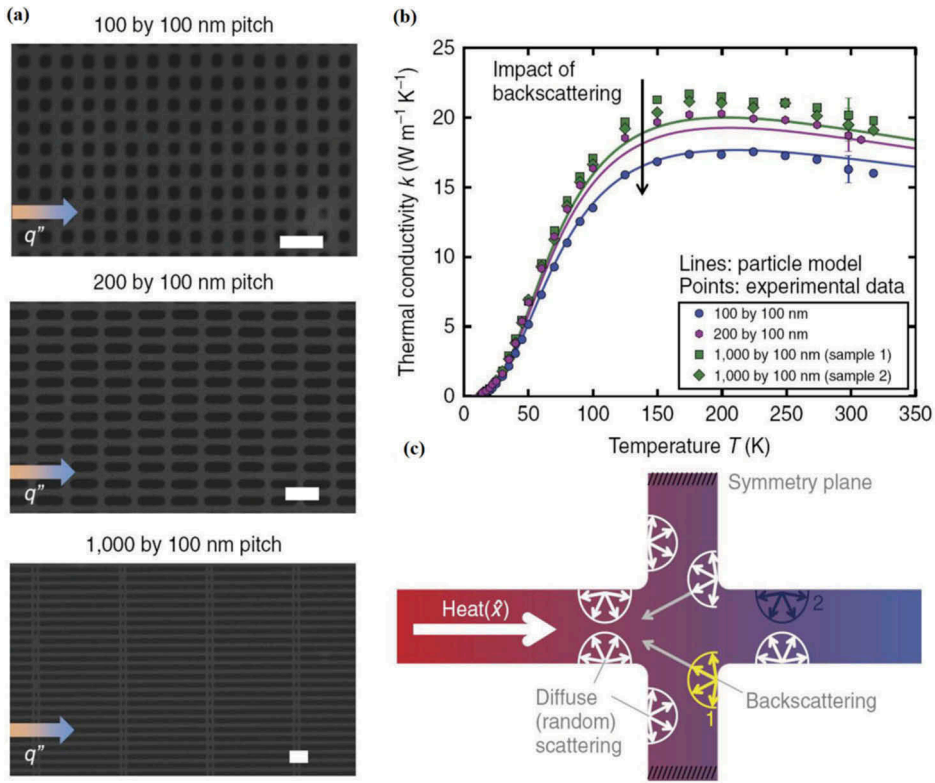

**Figure 10. Investigating backscattering effects with variable-pitch nanomeshes.** (a) SEM images of silicon NMs with varying pitch size along the direction of heat flux q". Scale bars, 200 nm. (b) Experimental results and particle model predictions for $\kappa(T)$ of four samples show that decreasing the pitch decreases $k$, as predicted by the backscattering effect. (c) Illustration of backscattering for diffuse surfaces.

Note: Adapted from Ref. [67].

measured thermal conductivity for the corresponding NMs is shown in Figure 10(b), which can be well explained by the backscattering mechanism. As shown in Figure 10(c), the neck intersection backscatters a larger percentage of incident phonons than the NW-like boundaries parallel to the global heat flux q", which leads to increased backscattering from short-pitch NMs. For example, 100% of the phonons are backscattered at point 1 (indicated in yellow), while the backscattering percentage is only 50% at point 2 (indicated in dark blue). Clearly, decreasing the aspect ratio $p_x/p_y$ increases the backscattering fraction, thereby thermal conductivity decreases as the aspect ratio $p_x/p_y$ decreases, as shown in Figure 10(b).

Wagner et al. fabricated PnCs with ordered and disordered lattices with equal filling fractions in free-standing Si membranes, and they found that the thermal conductivity of disordered PnCs was the same as that of ordered PnCs above room temperature [64]. Using ultrafast pump and probe spectrum and Raman thermometry, they found that even in ordered PnCs, the phonon coherence only existed in the range of less than 0.4 THz at room temperature, even for a hole wall roughness value as low as 1 nm [64]. From these recent experiments, it is suggested that the coherent phonons is not as important as expected on thermal transport in the regime where $\lambda = p = \Lambda_U$ and $\lambda \sim \delta$ (surface roughness). Very recently, on the basis of perturbation theory, Xie et al. presented a novel phonon-surface scattering mechanism from the perspective of bond order imperfections in the skin of nanostructures and interpreted the ultra-low thermal conductivity of 2D Si PnCs in the incoherent regime [169].

A new kind of metamaterial called phonon resonant structure has recently attracted considerable attention for blocking phonon transport [178–184]. In these nanostructures, such as pillar-based phononic crystal, a periodic array of nanopillars is built on the surface of NW or membrane. Standing waves can be generated inside the branches due to the total reflection of waves at the end of branches. As a result, a set of resonant frequencies can be obtained. Because of the band anticrossing, the resonances interact with the propagating modes and reduce their phonon group velocities. Xiong et al. performed molecular dynamics simulations on the thermal conductivity of branched SiNWs [178]. They found that low-frequency phonons can be easily manipulated with small resonators, which provides a powerful mechanism for blocking thermal transport.

Approaches based on wave interference allow control of nanoscale heat conduction by modifying physical transport properties that cannot be influenced by classical incoherent approaches, which attracts strong interest nowadays. At the present level of fabrication, the atomically smooth surfaces and short periodicity in NMs are still challenging to achieve, which limits the application of phonon coherence to tune thermal transport in PnCs at high temperature. However, further miniaturization, the use of materials with longer phonon MFP such as alloys and improvements in the hole fabrication process can enlarge the working temperature range. Very recently, Maire *et al.* demonstrated thermal conduction control by coherence in 2D Si PnCs in a large temperature range, until the transition to purely diffusive heat conduction was observed at 10 K [63]. It is reasonable to believe that advances in nanofabrication will keep broadening the working temperature range of phononic crystals, in the way that wave optics revolutionized the manipulations of light.

## 6. Direct detection of phonon coherence

Aforementioned phonon coherence has significant effect on thermal conductivity. In experiments, phonon coherence can be detected by measuring thermal conductivity. This is an indirect methodology. In this section, we introduce the direct method to detect phonon coherence.

The theory of quantum coherence defined in Equation (2) has been employed by Latour *et al.* to understand phonon coherence in SLs [185]. In this work, a phonon coherence function is defined based on Equation (2) between the velocity field of the crystal atoms at different space and time. Then, a frequency-dependent coherence length $l_c(\omega)$ can be defined such that $\frac{l_c(\omega)}{d_{SL}} \geq 1$ implies that phonon transport is coherent with $d_{SL}$ being the SL period.

Latour *et al.* first modeled argon SLs with Lennard-Jones (LJ) potential at 40 K [185]. The first layer has an LJ depth twice larger than the one of normal argon and the second layer has an LJ depth 2.5 larger than in the first one. Figure 11(a) shows the relationship between $\frac{l_c(\omega)}{d_{SL}}$ for different lattice periods of an argon SL where Log $\left(\frac{l_c(\omega)}{d_{SL}}\right) > 0$ implies that the transport is coherent. From the figure, it can be seen that increasing the lattice period causes a decrease in bandwidth of phonons being coherent. This explanation is consistent with experimental results in Figure 1. Figure 11(b) and (c) shows the size effect on coherence length. For a small SL period of 1 nm, the transport is coherent, while for SL with periodicity of 16 nm, the transport is incoherent and the coherence length no longer varies with system size. This trend is consistent with experimental results of length dependence in thermal conductivity in Figure 2.

Latour *et al.* then turned to silicon SL using Stillinger–Weber potential [185]. Figure 12 shows the coherence length for a silicon SL at different temperatures and the corresponding phonon DOS. As one can see, the low-frequency phonons are not affected much, but the high-frequency acoustics and optical phonons are greatly affected by temperature due to anharmonic scattering.

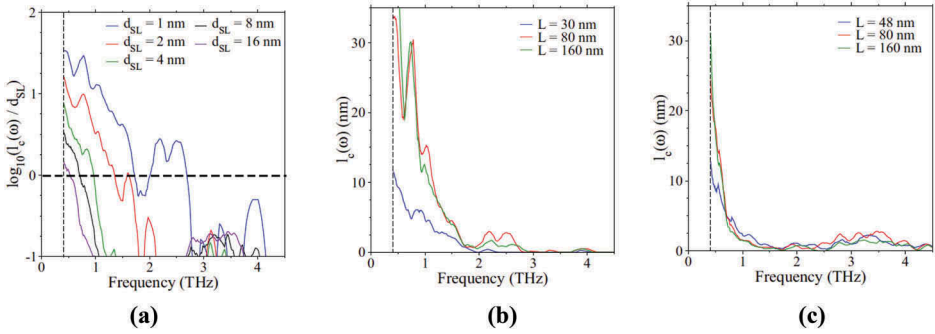

**(a)**                   **(b)**                   **(c)**

**Figure 11. Phonon coherence length versus frequency for different period and system size.** (a) Log $\left(\frac{l_c(\omega)}{d_{SL}}\right)$ is calculated to determine the phonon transport regime for an argon superlattice. When Log $\left(\frac{l_c(\omega)}{d_{SL}}\right) > 0$, phonon transport is coherent and vice versa. As the period size increases, the transport becomes more and more incoherent. (b) Effect of the system size on argon superlattices in the coherent regime with $d_{SL}$ = 1 nm. The coherence length decreases as the frequency increases. (c) Effect of the system size on argon superlattices in the incoherent regime with $d_{SL}$ = 16 nm. The coherence length is not affected as the system size increases.

Note: Adapted from Ref. [185].

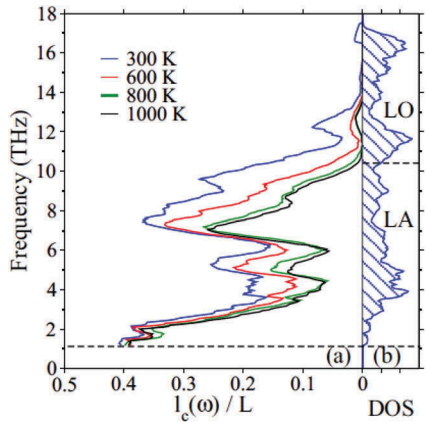

**Figure 12. Normalized spatial phonon coherence length for silicon superlattice at different temperatures** with $d_{SL}$ = 1 nm, for $T$ = 300, 600, 800 and 1000 K (a). (b) Phonon DOS of the silicon superlattice with $d_{SL}$ = 1 nm at 300 K.

Note: Adapted from Ref. [185].

Latour *et al.* also mapped out the different regimes of coherent transport relating four parameters, namely the coherence length $l_c$, SL period $d_{SL}$, the phonon MFP $\Lambda_{bulk1,2}$ of two bulk materials 1 and 2 in an SL and the system size $L$ [185]. Most notably in Figure 13(a) and (c), when coherence length $l_c$ is greater than the SL period $d_{SL}$, then phonon transport is coherent and thermal conductivity decreases with increasing

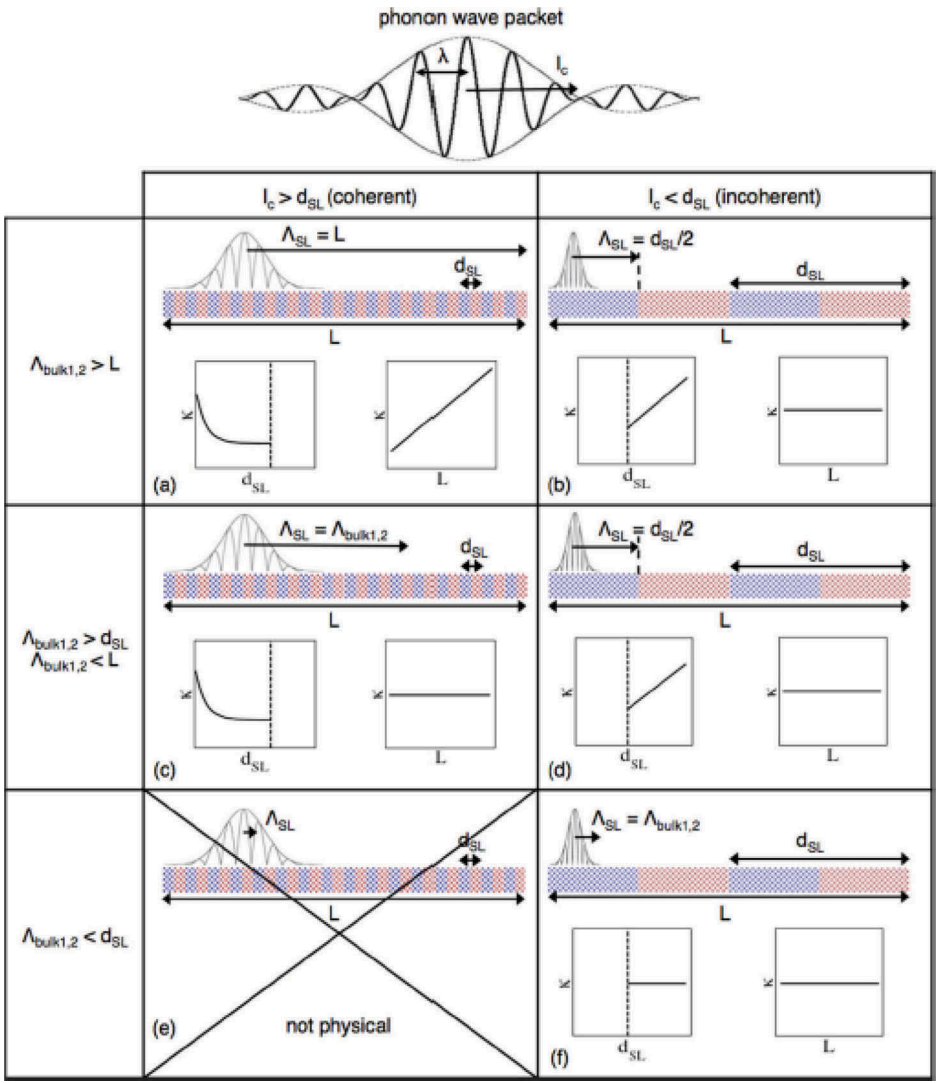

**Figure 13.** Schematic representation of all phonon characteristic lengths that are involved in the phonon transport in superlattices. $l_c$ is the phonon coherence length, $\lambda$ is the wavelength associated to the wave packet, $d_{SL}$ is the period thickness of the superlattice and $L$ is its length. For simplification, the two bulk materials 1 and 2 are assumed to have a similar mean free path, noted $\Lambda_{bulk1,2}$. Finally, $\Lambda_{SL}$ represents the mean free path of the wave packet in the superlattice. For each of these six cases, two trends for the thermal conductivity $\kappa$ are depicted: one as a function of the period thickness $d_{SL}$ with a constant length $L$ and one with respect to $L$ with a constant $d_{SL}$.

Note: Adapted from Ref. [185].

SL period. The turning of thermal conductivity happens when $l_c < d_{SL}$ in Figure 13(b), (d) and (f). Then, thermal conductivity becomes independent of system size and increases with SL period. The minimum point of thermal conductivity with interface density in Figure 1 is clearly explained with this effect.

Going beyond phonons in periodic structures, understanding phonon coherence using optical coherence theory will provide the ultimate evidence to describe phonons just like photons. There has been various work on generating and detecting coherent heat-carrying phonons in materials, especially at low temperatures [75–83], and it would be interesting to apply the concept of coherence in these experiments. Recently, work by Ding *et al.* further implemented concepts in quantum coherence in potentially characterizing phonon coherence [186,187]. In one work, Ding *et al.* proposed the use of two-photon interference such as coherent population trapping (CPT) or electromagnetic induced transparency (EIT) to detect phonon coherence [186].

The proposed scheme with CPT is shown in Figure 14(a). There are two photons of slightly different frequencies that are coherent with each other in a typical three-level configuration of a defect in a crystalline material. Phonons couple between levels $|3\rangle$ and $|1\rangle$, and $\Gamma_p$ is proportional to the incoherent phonon population while $W$ is proportional to the coherent phonon population. Figure 14(b) shows the ground state population levels $|3\rangle$ and $|1\rangle$ as a function of detuning $\delta_a$ while keeping $\delta_b = 0$ with and without incoherent phonons. It can be seen that incoherent phonons lead to a much-reduced transfer between the ground states on resonance $\delta_a = 0$. The excited state $|2\rangle$ population remains the same except at $\delta_a$ where the dip is sharply reduced. This dip is a characteristic feature of CPT and has been experimentally measured in defects of materials [188,189]. Incoherent phonons are treated as damping and are known to reduce the magnitude of this dip at resonance. However, when coherent phonons interact with the ground states $|3\rangle$ and $|1\rangle$, the excited state population behaves completely different from that in Figure 14(c). As shown in Figure 14(d), not only is the dip preserved on resonance, the line shape becomes asymmetric for different coherent phonon population. Ding *et al.* went further to discuss the physical reasoning behind this asymmetry and a similar discussion for EIT [186].

Last but not least, it is important to think of phonon coherence on a fundamental level. In quantum optics, the response of the photoelectric detector is proportional to the optical intensity and led to the coherence theory based on Equation (2). However, such a response may not be applicable for phonon detectors, and Equation (2) may not be valid for detecting phonon coherence. Ding *et al.* considered a popular phonon detector [187] called optical sideband detection [77,190,191]. Here, the relationship between the detector and the signal is no longer a simple function like that of the photoelectric detector. Ding *et al.* derived the second-order correlation for optical sideband detection and proposed an

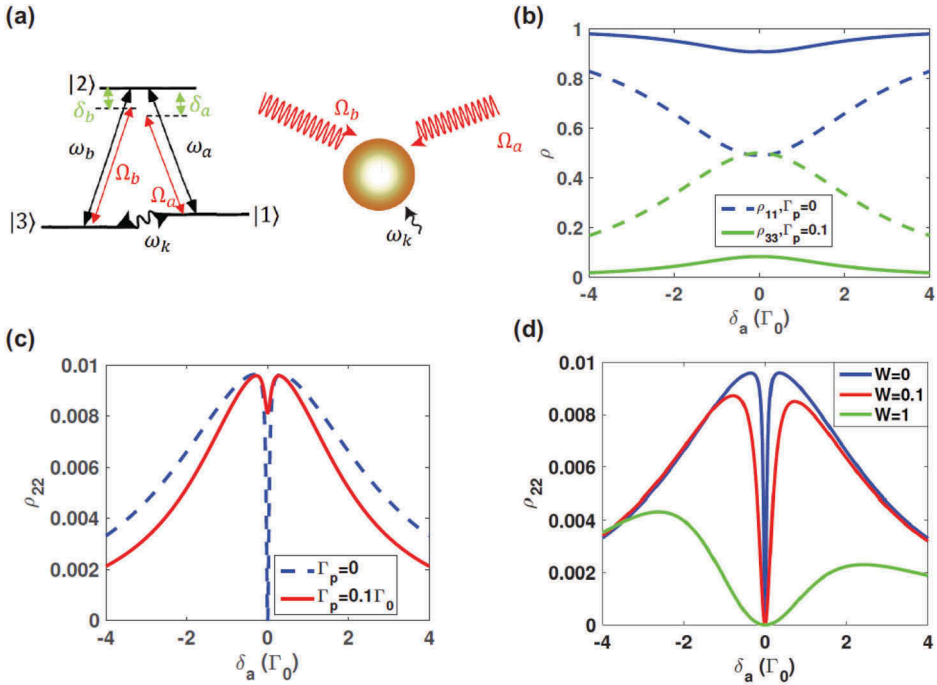

**Figure 14. Proposed method to detect coherent phonons with two-photon interference.** (a) Schematic of two-photon interference in a defect-based crystalline system. The emitter has a $\Lambda$-type energy level system. Levels $|1\rangle$ and $|3\rangle$ are part of the ground state manifold and the excited states $|2\rangle$ have a frequency of $\omega_a$ and $\omega_b$, respectively. The optical fields driving the $|2\rangle$-$|1\rangle$ transition and that driving $|2\rangle$-$|3\rangle$ transition have frequency $\Omega_a$ and $\Omega_b$ with detuning $\delta_a$ and $\delta_b$, respectively. (b) Populations in states $|1\rangle$ (blue lines) and $|3\rangle$ (green lines) as a function of the detuning $\delta_a$. The dashed line represents the case with no incoherent phonon. The solid line represents the case with incoherent phonon. (c) Population in level $|2\rangle$ with and without incoherent phonons represented by solid and dashed lines, respectively. (d) Population versus detuning $\delta_a$ in excited state $|2\rangle$ for different number of coherent phonons and no incoherent phonons.

Note: Adapted from Ref. [186].

interference experiment in Figure 15(a) which allows one to measure phonon coherence just like photons [187]. Figure 15(b) shows examples of the second-order correlation at zero delay for different phonon numbers. It can be seen that thermal and coherent phonons have different bounds. But unlike the case of $g_2(r_1, r_2, \tau) = 1$ discussed above for coherent electromagnetic fields, the bounds for $g_2(r_1, r_2, \tau)$ here depend on phonon number and other factors and can vary by orders of magnitude.

# 7. Conclusion

In this article, we took the reader on a tour presenting the state of the art of the topic termed 'phonon coherence' and its effect on thermal conductivity of various nanomaterials and nanostructures. We firstly discussed the classical

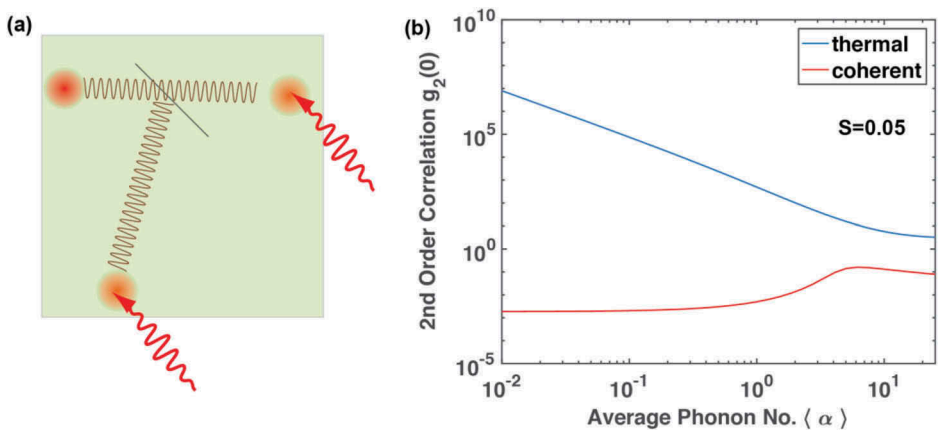

**Figure 15. Phonon coherence measurement with sideband fluorescence spectroscopy.** (a) Schematic of proposed measurement of phonon coherence using optical sideband detection. An optical excitation generates a phonon beam which gets split by a clean interface into two paths. (b) Phonon coherence for incoherent (thermal) and coherent phonons for a particular $S$ parameter which depends on the type of defects.

Note: Adapted from Ref. [187].

and quantum theory of coherence and the optical theory of classical and quantum coherence in heat transport. Then, we have reviewed recent experiments and theoretical works on thermal conductivity of SLs, NWs and NMs, emphasizing the effect of phonon coherence. Overall, phonon coherence can result in considerable reduction in thermal conductivity of NWs and phononic crystal, providing new concepts for improving the energy conversion efficiency in thermoelectric application. Finally, various theoretical methods to explore the phonon coherence using optical theory have been overviewed. Our intent is to give a state-of-the-art view with a balanced experimental and theoretical perspective. It is hoped that this review can provide important reference and guideline for the further systematic studies of this fundamental property of phonon, combining both experimental and theoretical efforts, which will be extremely helpful and greatly demanded to advance this field.

## Acknowledgments

GX is supported in part by the National Natural Science Foundation of China (Grant No. 11275163) and Natural Science Foundation of Hunan Province (Grant No. 2016JJ2131). DD and GZ gratefully acknowledge the financial support from the Agency for Science, Technology and Research (A*STAR), Singapore.

## Disclosure statement

No potential conflict of interest was reported by the authors.

## Funding

GX is supported in part by the National Natural Science Foundation of China [Grant No. 11275163] and Natural Science Foundation of Hunan Province [Grant No. 2016JJ2131]. DD and GZ gratefully acknowledge the financial support from the Agency for Science, Technology and Research (A*STAR), Singapore.